\documentclass[journal]{IEEEtran}

\usepackage{float}
\usepackage{caption}
\usepackage{subcaption}
\usepackage{longtable}
\usepackage{tabularx}
\usepackage{tabu}
\usepackage{multirow}
\usepackage{tikz}
\usepackage{graphicx}
\usepackage{tikzscale}
\usepackage{pgfplots}
\usepackage{setspace}
\usepackage{booktabs}
\usepackage{siunitx}
\usepackage{booktabs}
\usepackage{adjustbox}

\usepackage{algorithmic}
\usepackage{amsmath,amssymb,amsfonts}
\usepackage[ruled,linesnumbered]{algorithm2e}
\usepackage[font=small,labelfont=bf]{caption}
\SetKwFor{Foreach}{for each}{do}{endForEach}%
\SetKwFor{For}{for}{do}{endfor for}%
\SetKwIF{uIf}{ElseIf}{Else}{if}{then}{else if}{else}{endIf if}%
\SetKwFor{While}{while}{do}{endWhile while}%

\begin{document}

\title{
SMPC-based Federated Learning for 6G enabled Internet of Medical Things
}

\author{Aditya Pribadi Kalapaaking, Veronika Stephanie, Ibrahim Khalil, Mohammed Atiquzzaman, Xun Yi, Mahathir Almashor

}

\maketitle
\begin{abstract}
Rapidly developing intelligent healthcare systems are underpinned by Sixth Generation (6G) connectivity, ubiquitous Internet of Things (IoT), and Deep Learning (DL) techniques. This portends a future where 6G powers the Internet of Medical Things (IoMT) with seamless, large-scale, and real-time connectivity amongst entities. This article proposes a Convolutional Neural Network (CNN) based Federated Learning framework that combines Secure Multi-Party Computation (SMPC) based aggregation and Encrypted Inference methods, all within the context of 6G and IoMT. We consider multiple hospitals with clusters of mixed IoMT and edge devices that encrypt locally trained models. Subsequently, each hospital sends the encrypted local models for SMPC-based encrypted aggregation in the cloud, which generates the encrypted global model. Ultimately, the encrypted global model is returned to each edge server for more localized training, further improving model accuracy. Moreover, hospitals can perform encrypted inference on their edge servers or the cloud while maintaining data and model privacy. Multiple experiments were conducted with varying CNN models and datasets to evaluate the proposed framework's performance.
\end{abstract}


%
\IEEEpeerreviewmaketitle

\section{Introduction}\label{sec:intro}

Even at this nascent stage, the medical world is already looking beyond the significant breakthroughs in latency, mobility, and data rates present in 5G technologies \cite{yu2020deep}. With the rise of the Internet of Medical Things (IoMT), the research community has moved to 6G services as an enabler of ever more demanding AI-based diagnostics and e-healthcare services. Deep Learning (DL) methods have a voracious appetite for quality data and are integral to such next-generation medical services. This need for high-resolution and real-time imaging places stringent requirements on throughput and latency, which even 5G services will struggle to cope with in the coming decade \cite{tariq2020speculative}.

Potential applications within the healthcare realm include real-time robotic control and video-driven Human-Computer Interaction (HCI), allowing medical professionals to help patients remotely. IoMT relies on high inter-connectivity amongst devices, medical practitioners, and AI-based platforms. This enables healthcare organizations to improve the efficiency and quality of their services. Such inter-connectivity can drive accurate, and early diagnosis (and thus treatment) of various ailments, including computerized tomography (CT) scans for brain tumors and mammographies for breast cancers.

The increasing scale and variability of medical images challenge the capabilities of doctors. Contending the deluge of information can often lead to human errors and misdiagnoses. Accordingly, DL has shown great promise to aid medical practitioners by leveraging computer vision for image classification. Several studies have shown both accuracy and efficiency in medical diagnoses \cite{wang2020integrated}. Nevertheless, most IoMT devices are resource-constrained, which precludes their use for many DL tasks.

\begin{figure}[!h]
\centering
\includegraphics[width=1\linewidth]{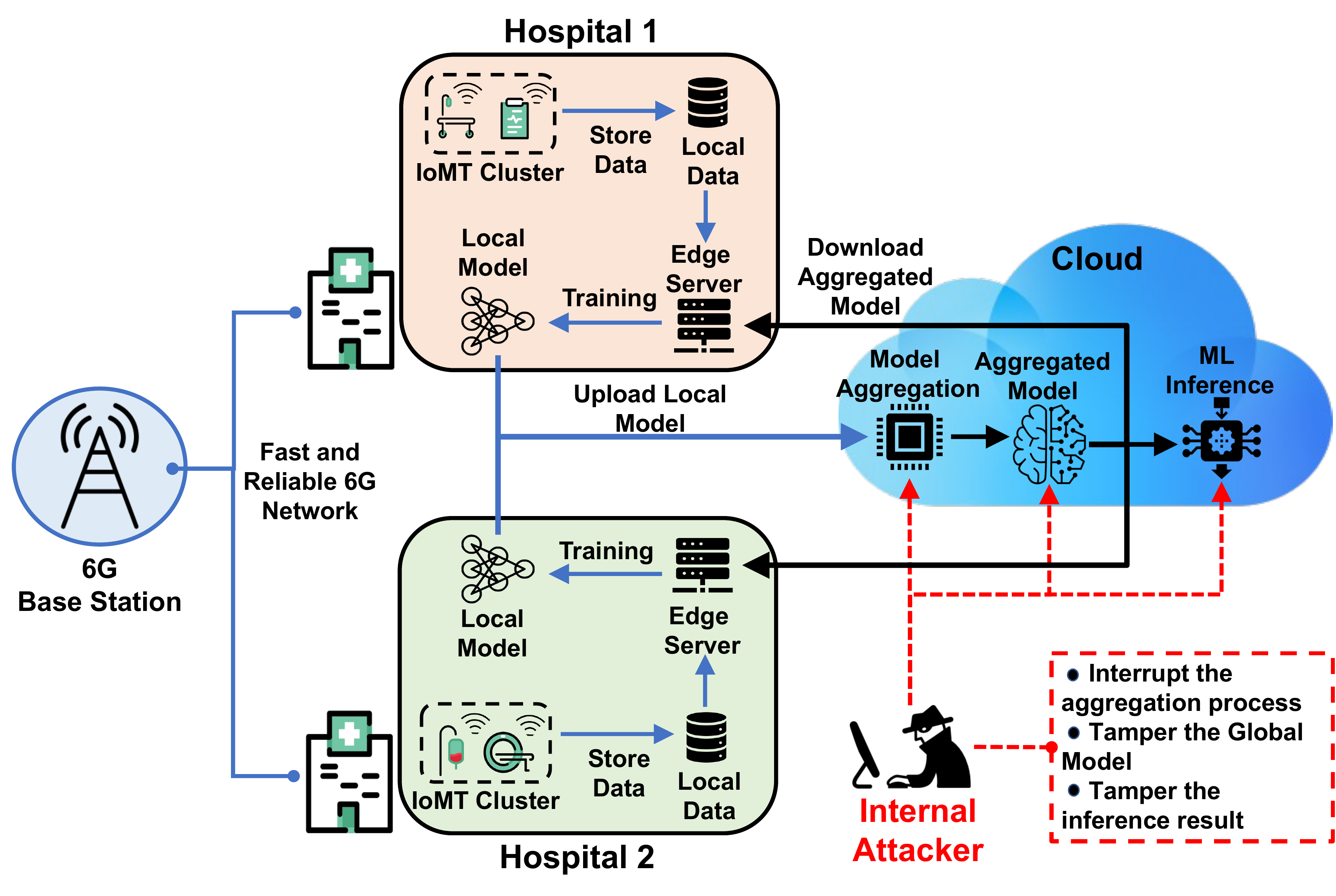}
\caption{\small{Traditional federated learning application and possible threat in healthcare scenario}}
\label{fig:scenario}
\end{figure}

Fig. \ref{fig:scenario} illustrates the cloud computing paradigm, which adds both compute and storage resources within the IoMT context. Here, the cloud is used to deploy the DL model for training and data inference. However, sending raw data from IoMT clusters to the cloud is intensive. Thus, edge computing servers are employed, which process the data before forwarding it to the cloud. 

Edge computing reduces the dependence of healthcare industries on centralized cloud infrastructure, which allows for a more responsive and agile IT network and, thus, more reliable patient services \cite{yu2020deep}. Additionally, with a 6G network, healthcare industries can enjoy near real-time information exchange within the IoMT ecosystem \cite{tariq2020speculative}.

It is known that high-performing DL models require extensive and diverse training datasets. This can only be obtained from multi-institutional or multi-national data sources and voluntary data sharing in the industry. While massive data collection is vital for DL, sharing sensitive patient data raises privacy concerns. Healthcare institutions may be legally prevented from sharing their data. Where sharing is possible, restrictions often apply, which results in incomplete or inadequate data.

Accordingly, \cite{9695218} proposed Federated Learning (FL) to allow parties to collaboratively train a model by sharing local updates with a parameter server. Intuitively, this method is safer than centralized training as public models learn from healthcare IoMT data without relying on a third-party cloud.

The application of the 6G network for existing FL methods allows IoMT systems to deliver more secure and reliable services. However, 6G presents many novel challenges, with few studies of it in IoMT and FL architectures. Furthermore, FL remains vulnerable to some attacks \cite{mothukuri2021survey}, which will be discussed further in section \ref{sec:issue}.

\textbf{Contributions:} We introduce a secure federated learning process that encrypts all local models and performs an SMPC-based model aggregation. Later the global model can be placed on the cloud to perform an encrypted inference process. Specifically, we:
\begin{enumerate}
    \item proposes an efficient architecture that leverages edge computing and 6G for IoMT applications.
    \item secures the 6G edge-computing federated learning model and the aggregation process, using SMPC.
    \item provide privacy by performing secure inference with multiple different parties in 6G network
\end{enumerate}

The rest of the article is organized as follows. Section \ref{sec:issue} discuss the possible threat in a current FL architecture. Section \ref{sec:related} discusses closely related work. The proposed framework is described in Section \ref{sec:framework}. Section \ref{sec:exp} shows experimental results and evaluates the performance of the proposed framework from different aspects. Finally, Section \ref{sec:con} concludes the framework with some concluding remarks.

\section{Security Issue}\label{sec:issue}

We present an FL-based IoMT-enabled healthcare system scenario to highlight current issues with traditional FL. Assume several smart hospitals that are geographically dispersed, with varying patient demographics and illnesses. Each hospital is equipped with a cluster of  IoMT devices (e.g., intelligent MRI and smart mammography).IoMT devices are used for monitoring patients' routines and detecting severe diseases. As they are resource-constrained and unable to perform typical ML processes, collected data is sent to the edge/cloud for ML training and validation. Hence, each hospital maintains an edge server with access to localized data. This helps to reduce external transmission costs and sensitive data exposure during ML training.

Limited access to larger and more complete datasets impacts the accuracy of local models trained on local data. Therefore, each hospital's edge server joins a cloud-based FL platform, where their locally trained models are aggregated with those from other hospitals. The aggregated models are called the global model. Eventually, the global model is sent back to each hospital. As a result, each hospital is benefited from increased accuracy and coverage.

However, the above approach suffers from the following security risks:
\begin{itemize}
    \item {\textit{Data privacy:} 
Sending local data to the cloud risks the privacy of said data. For example, a dishonest employee from the cloud service provider can act as an \textit{internal attacker} and collect private images or information (of the hospital or patients). The stolen data can be leaked or sold for personal gain, requiring a more robust and privacy-aware FL architecture.}
    \item {\textit{Local model and aggregation process privacy:} 
The aggregation process requires trained local models to be aggregated centrally, where they may be intercepted. Such local models are trained with sensitive data, and a bad actor (insider or external) can perform model inversion attacks before the aggregation process. Thus, a secure aggregation method is required to prevent such attacks.}
    \item {\textit{Result integrity when performing inference in the cloud:} In existing studies, IoMT devices send their data to the cloud for ML training and inference process. An attacker may access and tamper model results before being sent back to the hospitals. Hence, a secure inference method is required to ensure confidence in the integrity of results.}
\end{itemize}

\section{Related Work}\label{sec:related}

Several studies have presented federated learning in 6G architecture. Qu \textit{et al.} \cite{qu2021empowering} proposed an air unit or UAV as middlemen between cloud and end-users during the FL process. Using an air unit for model training and data inference process does not translate well to our healthcare scenario, as communication costs are high with inherent instability. Thus, we propose an FL architecture using intermediary edge servers between IoMT devices and the cloud. We also note that their architecture does not protect privacy in data and inter-party communications.

Zhou \textit{et al.} \cite{zhou2021two} proposed an efficient federated learning architecture for the Internet of Vehicles (IoV) using 6G. The work leverages edge servers to connect the cloud and vehicles. In the framework, a two-layer FL model aggregation is proposed. In the process, end-users send local models to edge servers. The first aggregation happens on the edge servers when all local models are received. This results in aggregated local models called \textit{aggregates}. The same edge servers leverage the cloud to perform a second aggregation upon completion to generate a global model called \textit{resultant}.

In preserving users' privacy, the framework first performs encryption on the local model parameters. This results in a high computation cost because each user's model needs to be encrypted after each training process. Furthermore, the secure aggregation process on the edge server violates users' privacy due to the decryption process done by the edge servers. In addition, the \textit{aggregates} produced by edge servers remain unencrypted as they are sent to the cloud. Similarly to the \textit{aggregates}, the \textit{resultant} produced by the cloud is left unencrypted.

Recently, several researchers have worked on a lightweight privacy-preserving method. Liu \textit{et al.} \cite{liu2021towards} proposed a secure and lightweight neural network inference system for secure intelligent medical diagnostic services. They enhance the additive secret sharing method to perform a lightweight inference process. Zheng \textit{et al.} \cite{9695218} built a lightweight encryption method based on cherrypicked cryptographic for aggregation method in federated learning. Their lightweight encryption method also has the ability to handle drop-out clients without exposing their secret keys. Wang \textit{et al.} \cite{9738843} developed a method to combat eavesdropping attacks in the wireless federated learning scenario. By sending jamming signals to the eavesdropper to enhance the server's secrecy throughput. However, the current privacy-preserving method only works to anticipate a specific attack. In contrast, we propose an SMPC-based cryptography method in the FL scenario that can secure both the aggregation and inference process.

\section{Proposed Framework}\label{sec:framework}
Here, we present our proposed general architecture in Section \ref{sec:sysArch}. Subsequently, the detailed architecture processes are discussed in three separate stages: Section \ref{sec:locModel} presents the local model generation; Section \ref{sec:secAgg} presents the SMPC-based encrypted aggregation protocol; and finally, Section \ref{sec:secEncIn} discusses the protocol used for SMPC-based encrypted inference.

\subsection{System Architecture}\label{sec:sysArch}
We propose an FL architecture that leverages SMPC to perform encrypted aggregation and encrypted inference.

\begin{figure}[!h]
\centering
\includegraphics[width=1\linewidth]{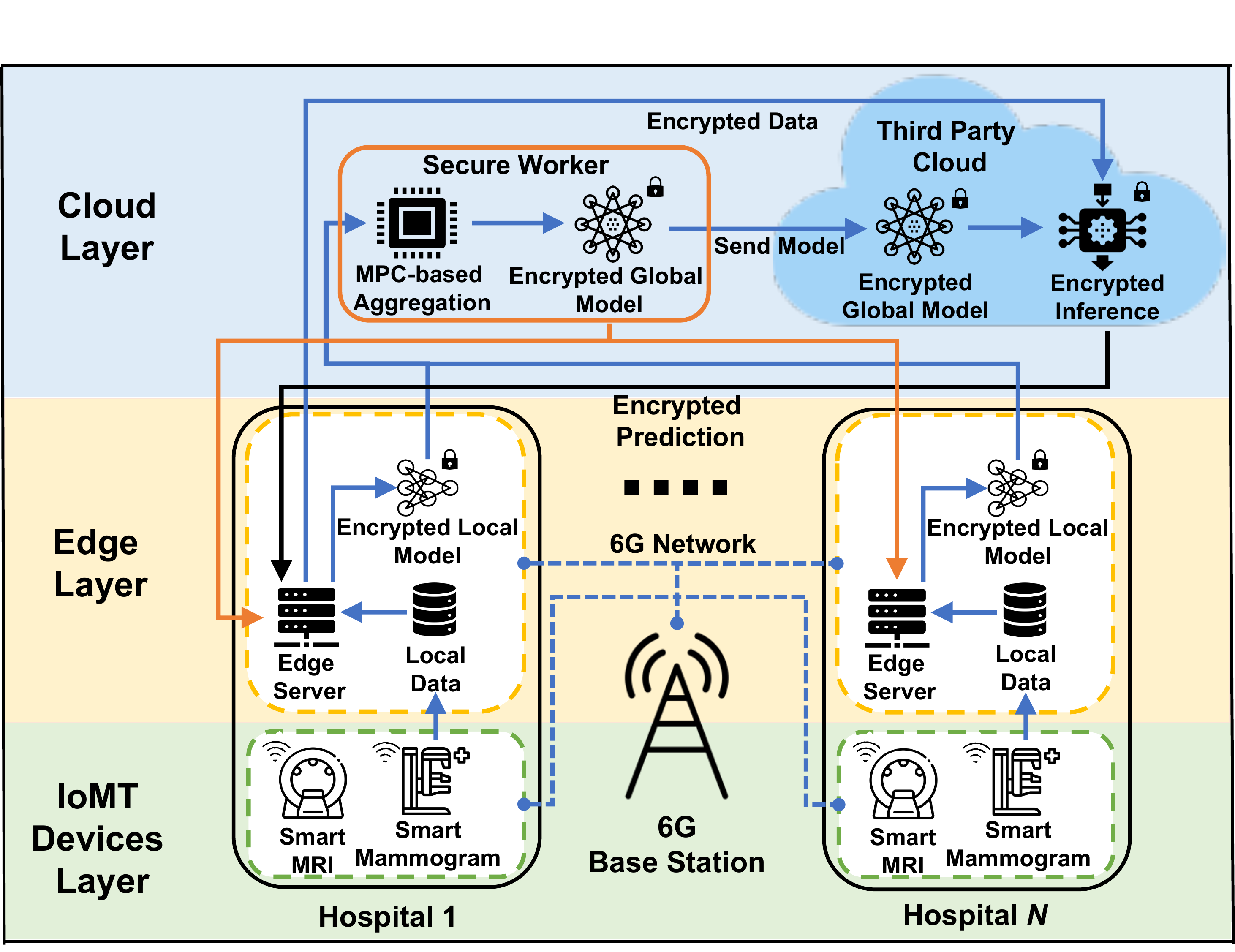}
\caption{\small{Overview of the proposed framework}}
\label{fig:architecture}
\end{figure}

As shown in Fig. \ref{fig:architecture}, our proposed architecture is divided into three layers, namely the IoMT devices layer, the edge layer, and the cloud layer. The IoMT devices layer comprises smart IoMT devices owned by hospitals. This layer is responsible for generating and transmitting data resources to the edge layer. The edge layer consists of edge servers owned by hospitals. This layer is responsible for executing edge computing tasks (e.g., local model training and local model encryption). The cloud layer is where the aggregation process of the encrypted local model happens. The resulting aggregated models are then stored in the cloud to be used for the encrypted inference process.

In our scenario, we assume that there exists $n$ number of hospitals $H$. Each hospital has an edge server $Z_n \in Z$ and IoMT devices $I = {i_1, i_2, ..., i_n}$ connected to $Z_n$. $Z_n$ collects data such as images from smart mammography or smart MRI produced by connected IoMT devices. A collection of $I$ connected to an edge server $Z_n$ creates a cluster of IoMT devices $C_{n} (1 \leq n \leq H)$. This scenario involves a multitude of devices and endpoints of varying capabilities and scale. Hence, a reliable and high-bandwidth connection is requisite for timely data transfers. This is a use case that 6G is well suited for.

Initially, the hospitals' edge servers receive images from IoMT devices to train their respective \textit{Local Models (LM)}. As mentioned, local model accuracy is impacted due to limited local data and resources. To overcome this, clusters of IoMT devices $C_{n}$ from multiple hospitals join the FL process. This allows multiple parties with varying datasets to contribute to the training process to produce diverse LMs. These LMs are then aggregated into a \textit{Global Model (GM)}, which would perform better than individual LMs. In our architecture, we perform encrypted aggregation on the LMs within the confines of a \textit{Trusted Third Party (TTP)} using an SMPC protocol. A secure multi-party computation or SMPC is a cryptographic technique that enables multiple parties to perform computation using their private data where no individual party can see the other parties’ data. The resulting encrypted GM is stored in the cloud. This allows participating hospitals to perform encrypted inference with high computational power.

\subsection{Local Model Generation}\label{sec:locModel}
In the local model generation, every hospital generates a local model based on its dataset collected from the IoMT cluster $C_{n}$. This step is similar to the initialization step in federated learning. Since $C_{n}$ contains many $I_{n}$, the communication time and the bandwidth used will be very high. We use 6G network to decrease the communication time and latency between each entity. 

In this scenario, we assume that every hospital has an edge server tasked with training local models. The DL approach used in our architecture is a Convolutional Neural Network (CNN) based image classification such as Resnet18 \cite{he2016deep}, and AlexNet \cite{krizhevsky2017imagenet}. In general, CNN image classification processes an input image and classifies it under certain categories of $t$ objects.

\begin{figure}[t]
\centering
\includegraphics[width=1.0\linewidth]{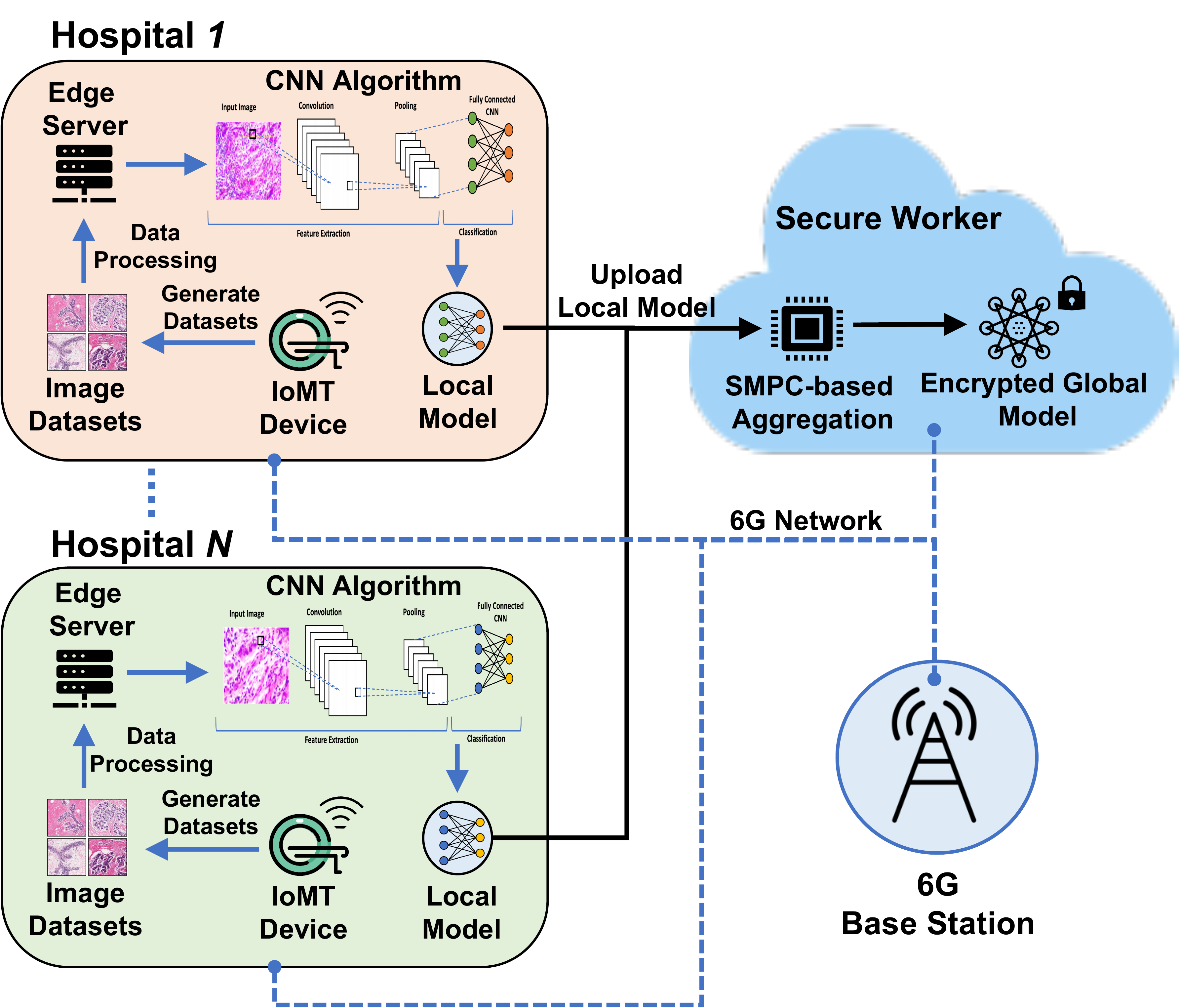}
\caption{\small{Local Model Generation}}
\label{fig:local_model_gen}
\end{figure}

In the local model training process, an edge server $Z_n$ of cluster $C_n$ recieves local image dataset $I_n$. The edge server process the input image as an array of pixels, depending on the image resolution. For example, an image with height $h$, width $w$, and dimension $d$, can be received as an array input of $h \times w \times d$. Note that in our scenario, $d$ is perceived as a color channel. After the input is processed, it is passed through different layers within the CNN models. The layers can include \textit{convolution layers with filters (Kernels)}, \textit{Pooling}, and \textit{fully connected layers (FC)}. Finally, CNN applies \textit{Softmax function} to classify an object with probabilistic values between $0$ and $1$. We refer to the trained model as \textit{local model} and denote it as $M_{L}$. An overview of the LM generation phase is illustrated in Fig. \ref{fig:local_model_gen}, where each hospital $H_n$ has an edge server $Z_n$ sends $M_{L}$ using a 6G network to the secure worker for the SMPC-based encrypted aggregation phase.

\subsection{SMPC-based Encrypted Model Aggregation}\label{sec:secAgg}

In our proposed scenario, we adopt Federated Averaging (FedAVG) \cite{mcmahan2017communication} algorithm as the base of our FL aggregation method. Without providing proper privacy protection, the FL application is not feasible. A local model produced by each FL participant is vulnerable to attacks such as model inversion and membership inference attacks. Hence, we leverage the SPDZ protocol to perform a secure aggregation process. SPDZ is the alias of the MPC protocol of Damg{\aa}rd \textit{et al}. \cite{10.1007/978-3-642-32009-5_38} for secure computation that allows parties to perform a computation on private values based on the additive secret sharing scheme. Additive secret sharing allows a trusted party $\mathcal{T}$ to share secrets $s_n \in s$ among $n$ parties $P_1, P_2, ... P_n$, such that to reveal $s$, $n$ participants must share their secrets. This process starts with a high number of prime number $Q$ generation. Then, $s$ is split into $n$ number of shares ${s_1, s_2, ..., s_n}$. In this scheme, the shares of $s$ must satisfy: $$s = \left(\sum_{i=1}^{n} s_i\right)\ mod\ Q$$ 
This protocol allows us to perform the sum of local model parameters without revealing each model's revealing value to other parties. Hence, we use this protocol to perform a secure aggregation process in our architecture.

In additive secret sharing, properties such as addition, subtraction, and multiplication are supported. In this particular paper, the SPDZ protocol for secure federated averaging employs the addition operation. Suppose that $P_1$ has a secret $s$ and $P_2$ has a secret $u$. Additionally, there exists a $\mathcal{T}$. In a particular case, $P_1$ and $P_2$ want to know the sum of their secrets without revealing the true value of their secrets. To calculate the sum, the protocol simply shares the secrets of the two parties into $n$ number of shares. 

Particularly, $s$ is divided into three shares $(s_1, s_2, s_3)$ and $u$ into another shares $(u_1, u_2, u_3)$. Each party holds one share of each secret. For example, $P_1$ holds ($s_1$, $u_1$), $P_2$ holds ($s_2$, $u_2$), and $\mathcal{T}$ holds ($s_3$, $u_3$). Then, to calculate the sum of the two secrets, each party adds up the shares that they hold using $m_n = (u_n + s_n)\ mod\ Q$, where $m_n$ is the sum of shares that party $n$ holds. Finally, the sum of shares are computed as follows: $$ u + s =  \left(\sum_{i=1}^{n} m_i\right)\ mod\ Q$$

\begin{figure}[t]
\centering
    \begin{subfigure}[b]{0,5\textwidth}
         \centering
        \includegraphics[width=1.0\linewidth]{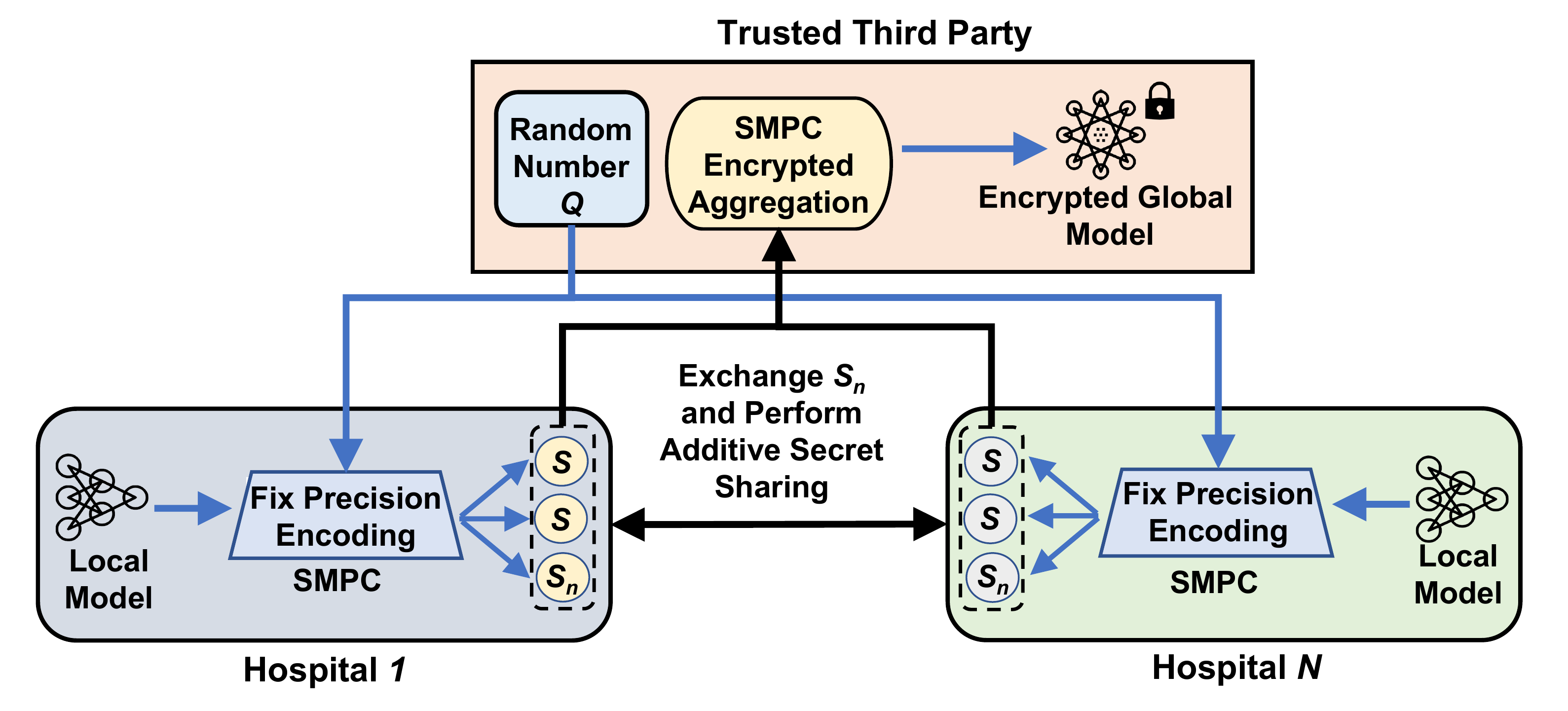}
        \caption{}
        \label{fig:enc_agg}
     \end{subfigure}
     \hfill
    \begin{subfigure}[b]{0.5\textwidth}
         \centering
        \includegraphics[width=1.0\linewidth]{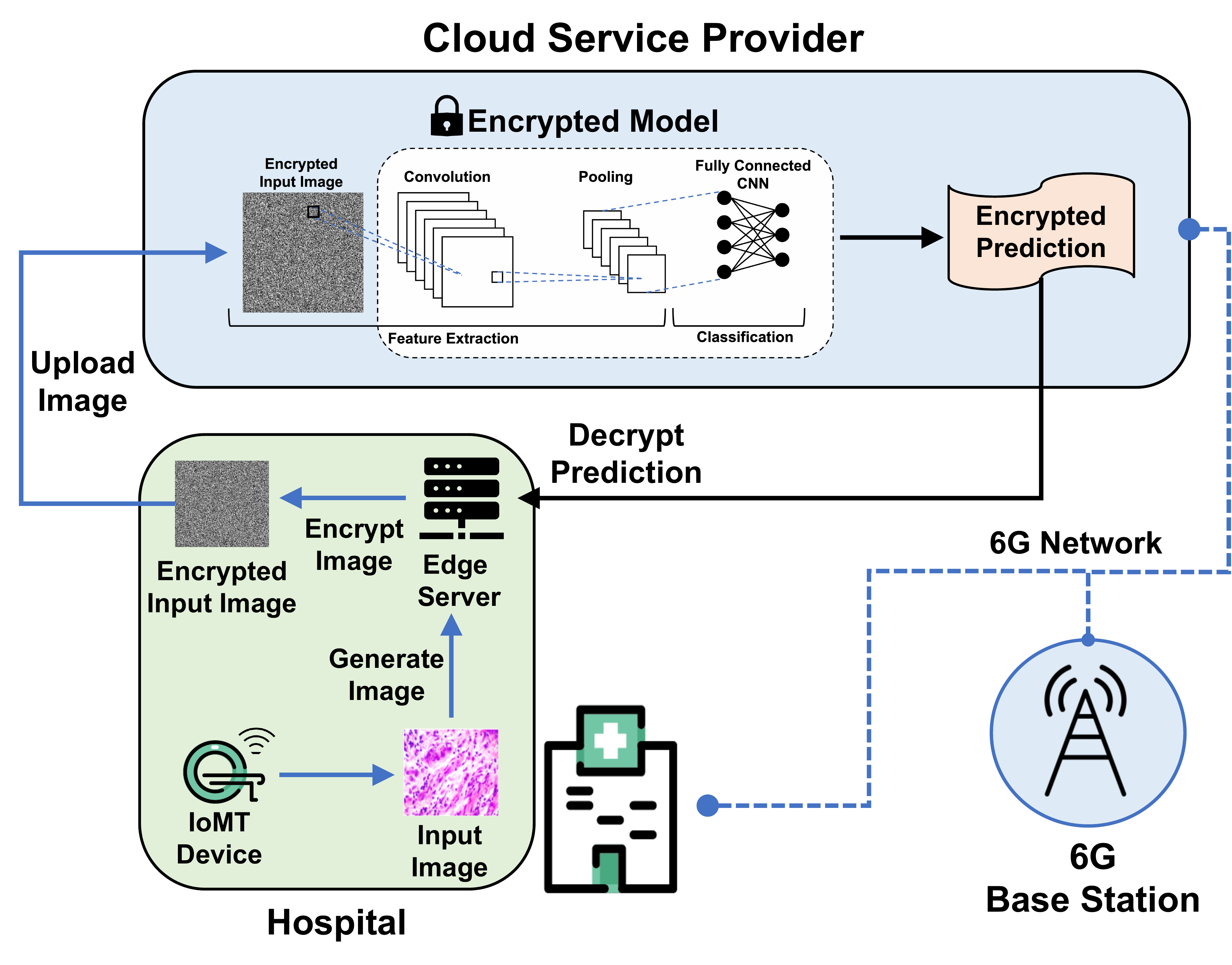}
        \caption{}
        \label{fig:enc_inf}
     \end{subfigure}
     \hfill
\caption{a) SMPC-based Encrypted Aggregation. b) SMPC-based Encrypted Inference}
\end{figure}

As communication between each party is crucial to produce an aggregated final model, the use of 6G is a critical way to ensure reliable inter-party connection in the secret-sharing process. The SMPC-based encrypted model aggregation are illustrated in Fig. \ref{fig:enc_agg}.

\subsection{SMPC-based Encrypted Inference}\label{sec:secEncIn}
In the architecture, a trusted party or secure worker $\mathcal{T}$ runs a secure aggregation process to produce a global model. The secure worker then encrypts this global model to ensure that no parties can perform model inversion or membership inference attacks. For this, we leverage additive secret sharing protocol and Function Secret Sharing (FSS) to allow data and model owners to keep their inputs and models confidential.

In the CNN setup, parameters and input data can be secretly shared amongst the party, and each operation in the process is treated independently. For addition operations in CNN layers, such as matrix multiplication layer, an additive secret sharing protocol mentioned in Section \ref{sec:secAgg} is used. However, for multiplication operation, Beaver triple, as mentioned in \cite{ziller2021pysyft}, is used as it fits with additive secret sharing.

For comparison operation in CNN layers such as \textit{Max Pooling}, Function Secret Sharing (FSS) is used. To achieve this, let there be a value $y$, and a comparison operation wants to know if $y \leq 0$. First, two shares $y_0$ and $y_1$ are generated from $y$. These shares are sent to the respective parties $P_0$ and $P_1$. Each party then masks these shares accordingly using masking values $a_{0}$ and $a_{1}$ by summing up their share and the mask values. Then, the summed values are added up together to produce a public value $x = y + a $. This value is then applied to FSS to obtain a shared output which determines if $y \leq 0$. In this case, a comparison function in FSS is defined as $$F(y + a) = F(x) = (x \leq a) = (y + a) \leq a = y \leq 0$$.

The 6G interplay in the encrypted inference phase allows an increase in encrypted prediction data download and upload rate. This enables parties $P_n$ or Hospitals to provide faster and more reliable services to the users. The SMPC-based encrypted inference are illustrated in Fig. \ref{fig:enc_inf}.

\section{Experimental Results}\label{sec:exp}
In this section, We provide discussions on the experimental setup, and dataset and model in Section \ref{sec:setup} and \ref{sec:data}, respectively. Section \ref{sec:results} shows experimental results and evaluations. We conducted several experiments to evaluate the performance of our proposed framework.

\subsection{Experimental Setup}\label{sec:setup}
We run the server and client sides in our experiments with AWS Sagemaker. We used \textit{ml.g4dn.16xlarge} series with 64 virtual CPUs, 256 GB Memory, and 1 NVIDIA T4 GPU. The G4DN series from AWS runs the application on a virtual CPU above the customized Intel Cascade Lake. It is optimized for machine learning inference and small-scale training. For the encrypted inference, we selected \textit{ml.p3.8xlarge} since it requires high computing power. This machine has 4 NVIDIA Tesla V100 with 64 GB memory and Peer to Peer connection between the GPU. In addition, it has 32 vCPUs and 244 GB of RAM. We built our federated learning application using PySyft \cite{ziller2021pysyft}.
\subsection{Dataset and Model}\label{sec:data}
We used the breast cancer dataset from Histopathological Database \cite{spanhol2016breast} to test the performance of our application in a practical IoMT scenario. It comprises 9,109 microscopic images of breast tumor tissue collected from 82 patients using different magnifying factors (40X, 100X, 200X, and 400X). The dataset is classified into 2,480  benign and 5,429 malignant samples (700X460 pixels, 3-channel RGB, 8-bit depth in each channel, PNG format). We partitioned the dataset evenly among all federated learning participants to train the local model. We evaluated our system using two different pre-trained machine learning models: Alexnet \cite{krizhevsky2017imagenet} and Resnet18 \cite{he2016deep}. The former has five convolutional layers and three fully connected layers with a total of 61 million parameters. The latter is a CNN comprising 18 layers and 11.7 million parameters.

\subsection{Results and Performance Evaluation}\label{sec:results}
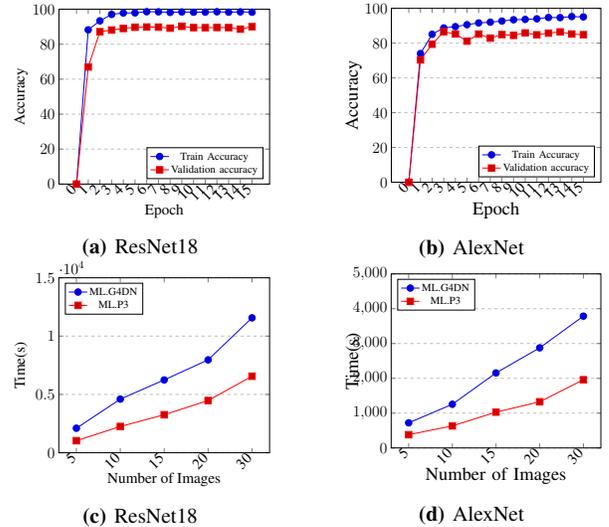
\begin{figure}[tbh!]
\centering
\begin{subfigure}[tbh!]{0.4\columnwidth}
    \resizebox{1\columnwidth}{!}{
            \begin{tikzpicture}
                \begin{axis}[
                    xlabel={Epoch},
                    ylabel={Accuracy},
                    symbolic x coords = {0,1,2,3,4,5,6,7,8,9,10,11,12,13,14,15},
                    xticklabel style={anchor= east,rotate=45 },
                    xtick=data,
                    ymax=100,
                    ymin=0,
                    legend pos=south east,
                    ymajorgrids=true,
                    grid style=dashed,
                    legend style={nodes={scale=1, transform shape}},
                    label style={font=\Large},
                    tick label style={font=\Large}
                ]
                \addplot+[mark size=3pt]
                    coordinates {
                    (0,0)
                    (1,88.23)
                    (2,93.36)
                    (3,97.05)
                    (4,97.86)
                    (5,97.94)
                    (6,98.60)
                    (7,98.53)
                    (8,98.19)
                    (9,98.46)
                    (10,98.30)
                    (11,98.35)
                    (12,98.57)
                    (13,98.33)
                    (14,98.50)
                    (15,98.33)
                    };
                \addplot+[mark size=3pt]
                    coordinates {
                    (0,0)
                    (1,67.01)
                    (2,87.15)
                    (3,88.11)
                    (4,89)
                    (5,89.71)
                    (6,89.83)
                    (7,89.77)
                    (8,89.19)
                    (9,90.22)
                    (10,89.51)
                    (11,89.45)
                    (12,89.58)
                    (13,89.51)
                    (14,88.62)
                    (15,90.03)
                    };
                \legend{Train Accuracy, Validation accuracy}
                \end{axis}
            \end{tikzpicture}
            }
    \caption{ResNet18}
    \label{exp_a}
\end{subfigure}
~
~
~
\begin{subfigure}[tbh!]{0.4\columnwidth}
    \resizebox{1\columnwidth}{!}{
        \begin{tikzpicture}
                \begin{axis}[
                    xlabel={Epoch},
                    ylabel={Accuracy},
                    symbolic x coords = {0,1,2,3,4,5,6,7,8,9,10,11,12,13,14,15},
                    xticklabel style={anchor= east,rotate=45 },
                    xtick=data,
                    ymax=100,
                    ymin=0,
                    legend pos=south east,
                    ymajorgrids=true,
                    grid style=dashed,
                    legend style={nodes={scale=1, transform shape}},
                    label style={font=\LARGE},
                    tick label style={font=\Large}
                ]
                \addplot+[mark size=3pt]
                    coordinates {
                    (0,0)
                    (1,74)
                    (2,85.07)
                    (3,88.73)
                    (4,89.47)
                    (5,90.61)
                    (6,91.63)
                    (7,92.04)
                    (8,92.69)
                    (9,93.44)
                    (10,93.63)
                    (11,93.98)
                    (12,94.74)
                    (13,94.67)
                    (14,95.27)
                    (15,95.08)
                    };
                \addplot+[mark size=3pt]
                    coordinates {
                    (0,0)
                    (1,70.33)
                    (2,79.35)
                    (3,86.57)
                    (4,85.23)
                    (5,81.14)
                    (6,85.17)
                    (7,82.86)
                    (8,84.91)
                    (9,84.46)
                    (10,85.87)
                    (11,84.78)
                    (12,85.74)
                    (13,86.45)
                    (14,85.23)
                    (15,84.85)
                    };
                \legend{Train Accuracy, Validation accuracy}
                \end{axis}
            \end{tikzpicture}
        }
    \caption{AlexNet}
    \label{exp_b}
\end{subfigure}
~
~
~
\begin{subfigure}[tbh!]{0.4\columnwidth}
    \resizebox{1\columnwidth}{!}{
            \begin{tikzpicture}
                \begin{axis}[
                    xlabel={Number of Images},
                    ylabel={Time(s)},
                    symbolic x coords = {5,10,15,20,30},
                    xticklabel style={anchor= east,rotate=45 },
                    xtick=data,
                    ymax=15000,
                    ymin=0,
                    legend pos=north west,
                    ymajorgrids=true,
                    grid style=dashed,
                    legend style={nodes={scale=1, transform shape}},
                    label style={font=\Large},
                    tick label style={font=\Large}
                ]
                \addplot+[mark size=3pt]
                    coordinates {
                    (5,2100)
                    (10,4597)
                    (15,6243)
                    (20,7965)
                    (30,11565)
                    };
                \addplot+[mark size=3pt]
                    coordinates {
                    (5,1030)
                    (10,2249)
                    (15,3256)
                    (20,4475)
                    (30,6558)
                    };
                \legend{ML.G4DN, ML.P3}
                \end{axis}
            \end{tikzpicture}
            }
    \caption{ResNet18}
    \label{exp_c}
\end{subfigure}
~
~
~
\begin{subfigure}[tbh!]{0.4\columnwidth}
    \resizebox{1\columnwidth}{!}{
        \begin{tikzpicture}
                \begin{axis}[
                    xlabel={Number of Images},
                    ylabel={Time(s)},
                    symbolic x coords = {5,10,15,20,30},
                    xticklabel style={anchor= east,rotate=45 },
                    xtick=data,
                    ymax=5000,
                    ymin=0,
                    legend pos=north west,
                    ymajorgrids=true,
                    grid style=dashed,
                    legend style={nodes={scale=1, transform shape}},
                    label style={font=\LARGE},
                    tick label style={font=\Large}
                ]
                \addplot+[mark size=3pt]
                    coordinates {
                    (5,721)
                    (10,1254)
                    (15,2155)
                    (20,2876)
                    (30,3786)
                    };
                \addplot+[mark size=3pt]
                    coordinates {
                    (5,380)
                    (10,633)
                    (15,1028)
                    (20,1326)
                    (30,1959)
                    };
                \legend{ML.G4DN, ML.P3}
                \end{axis}
            \end{tikzpicture}
        }
    \caption{AlexNet}
    \label{exp_d}
\end{subfigure}
\caption{Federated learning accuracy with encrypted aggregation using a)ResNet18 Model; b)AlexNet Model; Encrypted Inference time cost for different number of images with a)ResNet18 Model; b)AlexNet Model;}
\label{exp1}
\end{figure}

We first evaluate our architecture by checking the accuracy of the transfer learning. In this experiment, we test federated learning with an encrypted aggregation setup against the breast cancer dataset. As shown in Fig. \ref{exp_a} and Fig. \ref{exp_b} at the beginning, they start from 0\% and goes up to 84\% accuracy during the first epoch. This kind of result can be achieved when we are using transfer learning.

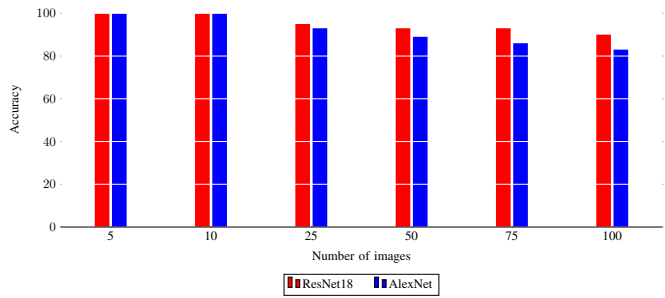
\begin{figure}[tbh!]
    \centering
    \resizebox{1\columnwidth}{!}{
        \begin{tikzpicture}
            \centering
            \begin{axis}[
                ybar, axis on top,
                height=8cm, width=18cm,
                bar width=0.4cm,
                ymajorgrids, tick align=inside,
                major grid style={draw=white},
                enlarge y limits={value=.1,upper},
                ymin=0,
                ymax = 100,
                axis x line*=bottom,
                y axis line style={opacity=0},
                tickwidth=0pt,
                enlarge x limits=true,
                legend style={
                    at={(0.5,-0.2)},
                    anchor=north,
                    legend columns=-1,
                    /tikz/every even column/.append style={column sep=0.5cm}
                },
                ylabel={Accuracy},
                xlabel={Number of images},
                symbolic x coords={5,10,25,50,75,100},
                xtick=data,
                nodes near coords,
                every node near coord/.append style={rotate=90, anchor=west},
                nodes near coords={}
            ]
            \addplot [draw=none, fill=red] coordinates {
                (5,100)
                (10,100)
                (25,95)
                (50,93)
                (75,93)
                (100,90)
            };
            \addplot [draw=none, fill=blue] coordinates {
                (5,100)
                (10,100)
                (25,93)
                (50,89)
                (75,86)
                (100,83)
            };
            \legend{ResNet18, AlexNet}
            \end{axis}
        \end{tikzpicture}
    }
    \caption{Encrypted inference accuracy with different number of images using ResNet and AlexNet}
    \label{exp2}
\end{figure}

In a practical healthcare scenario, transfer learning can be helpful because not all hospitals can generate their model as there is a need for IoMT devices to generate the data, in addition to powerful computing resources. In Fig. \ref{exp_a} and Fig. \ref{exp_b} we also perform experiments with encrypted aggregation process using SMPC. As can be seen, we obtained high training and evaluation accuracies. Specifically, we achieved up to 98\% for Resnet during training, and up to 90\% during validation. Similarly for AlexNet, the accuracies are up to 95\% and 85\% respectively, for training and validation.
In Fig. \ref{exp2} we evaluate the encrypted inference with differing image batch sizes. We run each batch of images five times for the encrypted inference and average the overall result. In our scenario, we send encrypted data to the cloud, so both the data and the inference are performed in an encrypted way. Starting with 5 and 10 images, both ResNet18 and AlexNet achieve 100\% accuracy when completing the encrypted inference. When the image batch size reaches 25, the accuracy from both models slightly decreases while performing the encrypted inference; However, it is still above 90\%. From the overall experiment results for the accuracy with encrypted inference and data, the fewer images we performed in the encrypted inference, the more accurate the outcome is. The plot shows that the ResNet18 model has higher accuracy than the AlexNet model for every different image batch size.

In the experiment shown in Fig. \ref{exp_c} and Fig. \ref{exp_d}, we evaluate the encrypted inference time cost to see the performance of our system with a different number of image batch sizes. In this scenario, we will assess the ResNet18 and AlexNet machine learning model with two other VMs, the \textit{ML.G4DN} is the machine we assume every hospital will have, and \textit{ML.P3} is the cloud that has powerful computation power with multiple GPU.

As is seen, both ResNet18 and AlexNet have linearly increased for the time cost. Performing encrypted inference with ResNet18 will take a longer time to compare to AlexNet. However, from the previous experiment shown in Fig. \ref{exp2}, the accuracy result from ResNet18 is stable above 90\%. From both experiments result in Fig. \ref{exp_c} and Fig. \ref{exp_d} we also see similar results, while performing encrypted inference with the more capable computing power availble reduces the time cost by up to 50\%. From this result, we see that even though each hospital has the encrypted global model, their limited resources cannot handle large image batch sizes which leads to a stronger cloud computational requirement.

\section{Future Research Direction}\label{sec:fut}
In this article, SMPC-based encrypted aggregation and secure inference have been introduced to improve the security of 6G federated learning architectures. However, there remain several challenges requiring further research:

\textbf{6G Infrastructure.} 
The communication bottleneck is considered as a major challenge in an IoT-based federated learning environment. IoT devices communicate using wireless networks. With additional security settings that require multiple communication rounds, such as SMPC, huge internet bandwidth is imperative. The advanced development of the 6G network can solve the limited bandwidth issue. However, for such frameworks to be accepted, 6G infrastructure must be established across continents. Due to the high costs of deploying 6G infrastructures, there needs to be better research into expanding 6G adoption strategies.

\textbf{Lightweight Machine Learning.} 
The resource-limited nature of IoT devices is a barrier to on-device training of DL models. Accordingly, reducing on-device model complexity (by outsourcing training to edge servers) while maintaining privacy protections is an ongoing concern.

\textbf{Novel Anomaly Detection Mechanism.} 
There is a need to identify and prevent malicious IoT devices from sending data and jeopardizing FL training. Further research to detect system anomalies is encouraged, such as metric evaluation from IoT devices to identify malicious nodes.

\textbf{Cost-efficient communication.} 
As the number of IoT devices participating in FL increases, communication cost remains a crucial barrier to wide adoption within the 6G context. A cost-efficient mechanism must be developed to make the FL model acceptable for 6G networks with extremely heterogeneous devices and network delays.

\section{Conclusion}\label{sec:con}
This paper proposes privacy-preserving federated learning with SMPC-based encrypted model aggregation and inference for a 6G-based IoMT environment. The main objective is to ensure local models from each hospital remain private during the aggregation process and to perform secure inference in third-party cloud platforms.

The model from each hospital is encrypted and split into several shares. Later, the secret shares from each party are computed using additive secret sharing to generate a global model. Eventually, the encrypted global model is sent to a hospital-controlled edge server for an iterative FL process and stored in the cloud to perform an encrypted inference.

In our architecture, we also leverage edge computing, and thanks to the 6G network, the IoMT devices can maintain a stable, timely, and reliable connection with hospital-owned edge servers. We conducted several experiments that show the proposed framework has high accuracy despite performing encrypted aggregation. In addition, we have noted several critical challenges while performing secure multi-party computation and federated learning processes in future research directions. In the future, we plan to develop a more lightweight encrypted inference mechanism that can be done on edge servers.

\section*{Acknowledgement}
This work is supported by the Australian Research Council Discovery Project (DP210102761).

\bibliographystyle{IEEEtran}
\bibliography{References}

\begin{thebibliography}{10}
\providecommand{\url}[1]{#1}
\csname url@samestyle\endcsname
\providecommand{\newblock}{\relax}
\providecommand{\bibinfo}[2]{#2}
\providecommand{\BIBentrySTDinterwordspacing}{\spaceskip=0pt\relax}
\providecommand{\BIBentryALTinterwordstretchfactor}{4}
\providecommand{\BIBentryALTinterwordspacing}{\spaceskip=\fontdimen2\font plus
\BIBentryALTinterwordstretchfactor\fontdimen3\font minus
  \fontdimen4\font\relax}
\providecommand{\BIBforeignlanguage}[2]{{%
\expandafter\ifx\csname l@#1\endcsname\relax
\typeout{** WARNING: IEEEtran.bst: No hyphenation pattern has been}%
\typeout{** loaded for the language `#1'. Using the pattern for}%
\typeout{** the default language instead.}%
\else
\language=\csname l@#1\endcsname
\fi
#2}}
\providecommand{\BIBdecl}{\relax}
\BIBdecl

\bibitem{yu2020deep}
S.~Yu, X.~Chen, Z.~Zhou, X.~Gong, and D.~Wu, ``{When Deep Reinforcement
  Learning Meets Federated Learning: Intelligent Multitimescale Resource
  Management for Multiaccess Edge Computing in 5G Ultradense Network},''
  \emph{IEEE Internet of Things Journal}, vol.~8, no.~4, pp. 2238--2251, 2020.

\bibitem{tariq2020speculative}
F.~Tariq, M.~R. Khandaker, K.-K. Wong, M.~A. Imran, M.~Bennis, and M.~Debbah,
  ``{A speculative study on 6G},'' \emph{IEEE Wireless Communications},
  vol.~27, no.~4, pp. 118--125, 2020.

\bibitem{wang2020integrated}
W.~Wang, F.~Liu, X.~Zhi, T.~Zhang, and C.~Huang, ``{An Integrated deep learning
  algorithm for detecting lung nodules with low-dose CT and its application in
  6G-enabled internet of medical things},'' \emph{IEEE Internet of Things
  Journal}, vol.~8, no.~7, pp. 5274--5284, 2020.

\bibitem{9695218}
Y.~Zheng, S.~Lai, Y.~Liu, X.~Yuan, X.~Yi, and C.~Wang, ``{Aggregation Service
  for Federated Learning: An Efficient, Secure, and More Resilient
  Realization},'' \emph{IEEE Transactions on Dependable and Secure Computing},
  pp. 1--1, 2022.

\bibitem{mothukuri2021survey}
V.~Mothukuri, R.~M. Parizi, S.~Pouriyeh, Y.~Huang, A.~Dehghantanha, and
  G.~Srivastava, ``{A survey on security and privacy of federated learning},''
  \emph{Future Generation Computer Systems}, vol. 115, pp. 619--640, 2021.

\bibitem{qu2021empowering}
Y.~Qu, C.~Dong, J.~Zheng, H.~Dai, F.~Wu, S.~Guo, and A.~Anpalagan,
  ``{Empowering Edge Intelligence by Air-Ground Integrated Federated
  Learning},'' \emph{IEEE Network}, vol.~35, no.~5, pp. 34--41, 2021.

\bibitem{zhou2021two}
X.~Zhou, W.~Liang, J.~She, Z.~Yan, and K.~Wang, ``{Two-layer Federated Learning
  with Heterogeneous Model Aggregation for 6G Supported Internet of
  Vehicles},'' \emph{IEEE Transactions on Vehicular Technology}, 2021.

\bibitem{liu2021towards}
X.~Liu, Y.~Zheng, X.~Yuan, and X.~Yi, ``{: Towards Secure and Lightweight Deep
  Learning as a Medical Diagnostic Service},'' in \emph{European Symposium on
  Research in Computer Security}.\hskip 1em plus 0.5em minus 0.4em\relax
  Springer, 2021, pp. 519--541.

\bibitem{9738843}
T.~Wang, Y.~Li, Y.~Wu, and T.~Q. Quek, ``{Secrecy driven Federated Learning via
  Cooperative Jamming: An Approach of Latency Minimization},'' \emph{IEEE
  Transactions on Emerging Topics in Computing}, pp. 1--1, 2022.

\bibitem{he2016deep}
K.~He, X.~Zhang, S.~Ren, and J.~Sun, ``{Deep residual learning for image
  recognition},'' in \emph{Proceedings of the IEEE conference on computer
  vision and pattern recognition}, 2016, pp. 770--778.

\bibitem{krizhevsky2017imagenet}
A.~Krizhevsky, I.~Sutskever, and G.~E. Hinton, ``{ImageNet classification with
  deep convolutional neural networks},'' \emph{Communications of the ACM},
  vol.~60, no.~6, pp. 84--90, 2017.

\bibitem{mcmahan2017communication}
B.~McMahan, E.~Moore, D.~Ramage, S.~Hampson, and B.~A. y~Arcas,
  ``{Communication-efficient learning of deep networks from decentralized
  data},'' in \emph{Artificial intelligence and statistics}.\hskip 1em plus
  0.5em minus 0.4em\relax PMLR, 2017, pp. 1273--1282.

\bibitem{10.1007/978-3-642-32009-5_38}
I.~Damg{\aa}rd, V.~Pastro, N.~Smart, and S.~Zakarias, ``Multiparty computation
  from somewhat homomorphic encryption,'' in \emph{Advances in Cryptology --
  CRYPTO 2012}, R.~Safavi-Naini and R.~Canetti, Eds.\hskip 1em plus 0.5em minus
  0.4em\relax Berlin, Heidelberg: Springer Berlin Heidelberg, 2012, pp.
  643--662.

\bibitem{ziller2021pysyft}
A.~Ziller, A.~Trask, A.~Lopardo, B.~Szymkow, B.~Wagner, E.~Bluemke, J.-M.
  Nounahon, J.~Passerat-Palmbach, K.~Prakash, N.~Rose \emph{et~al.}, ``{PySyft:
  A Library for Easy Federated Learning},'' in \emph{Federated Learning
  Systems}.\hskip 1em plus 0.5em minus 0.4em\relax Springer, 2021, pp.
  111--139.

\bibitem{spanhol2016breast}
F.~A. Spanhol, L.~S. Oliveira, C.~Petitjean, and L.~Heutte, ``{Breast cancer
  histopathological image classification using convolutional neural
  networks},'' in \emph{2016 international joint conference on neural networks
  (IJCNN)}.\hskip 1em plus 0.5em minus 0.4em\relax IEEE, 2016, pp. 2560--2567.

\end{thebibliography}

\end{document}